# Lattice stability and formation energies of intrinsic defects in $Mg_2Si$ and $Mg_2Ge$ via first principles simulations

Philippe Jund[a], Romain Viennois[a], Catherine Colinet[b], Gilles Hug[c], Mathieu Fèvre[c] and Jean-Claude Tédenac[a]

[a] Institut de Chimie Moléculaire et des Matériaux I.C.G., UMR-CNRS 5253, Université Montpellier II, Place E. Bataillon, 34095 Montpellier Cedex 5, France
[b] Science et Ingénierie des Matériaux et Procédés, Grenoble INP, UJF, CNRS, 38402 Saint Martin d'Hères Cedex
[c] Laboratoire d'Etude des Microstructures, UMR 104 ONERA-CNRS ONERA, Boîte Postale 72, 92322 Châtillon Cedex, France

Abstract

We report an *ab initio* study of the semiconducting $Mg_2X$ (with X = Si, Ge) compounds and in particular we analyze the formation energy of the different point defects with the aim to understand the intrinsic doping mechanisms. We find that the formation energy of $Mg_2Ge$ is 50 % larger than the one of $Mg_2Si$, in agreement with the experimental tendency. From the study of the stability and the electronic properties of the most stable defects taking into account the growth conditions, we show that the main reason for the n-doping in these materials comes from interstitial magnesium defects. Conversely, since other defects acting like acceptors such as Mg vacancies or multivacancies are more stable in $Mg_2Ge$ than in $Mg_2Si$, this explains why $Mg_2Ge$ can be of n or p type, contrary to $Mg_2Si$. The finding that the most stable defects are different in $Mg_2Si$ and $Mg_2Ge$ and depend on the growth conditions is important and must be taken into account in the search of the optimal doping to improve the thermoelectric properties of these materials.

Keywords: thermoelectric materials, silicides, germanides, ab initio simulations, defects

## 1. Introduction

Since the nineties, new thermoelectric materials and new concepts to improve the thermoelectric materials have been developed. These improvements together with the development of new synthesis techniques and the urgent need to find new green energy sources have permitted to focus the research on thermoelectricity during the last years. A key point is to find not only materials with a dimensionless figure of merit ZT larger than 1, in order to have a competitive efficiency for thermogeneration of electricity, but also to develop environment-friendly materials made of cheap, abundant, weakly toxic and recyclable elements. Silicide compounds fulfill these requirements and are very promising candidates for high temperature applications [1]. In particular $Mg_2Si$ has been the topic of numerous experimental [2] and numerical investigations [3]. $Mg_2Si$ crystallizes in the cubic anti-fluorite structure under ambient conditions with a lattice parameter in the range 6.34-6.39 Å [1,4]. The unit cell consists of four silicon and eight magnesium atoms. The silicon atom occupies the 4a Wyckoff site and the magnesium atoms occupy the 8c (0.75, 0.25, 0.25) sites, respectively (see Figure 1). The four interstitial sites correspond to the 4b Wyckoff positions. The Fm-3m space group fixes the fractional coordinates of all atoms. $Mg_2Si$ is an *n*-type semiconductor with an indirect band gap of 0.66-0.78 eV while $Mg_2Ge$ is a semiconductor with a slightly smaller indirect band gap of 0.57-0.74 eV [5]. Despite the numerous studies dedicated to $Mg_2Si$, some discrepancies between experimental and simulation results exist especially concerning the existence or not of several different high pressure phases [6-9] and the formation enthalpies of both the pure compound [10-15] and the intrinsic defects (vacancies, antisites) [14-15].

To improve its thermoelectric properties, $Mg_2Si$ has often been alloyed with germanium or/and tin using different synthesis conditions. Whatever these conditions are, it systematically results in an n-type intrinsic doping [16-18]. In order to explain the difficulty to synthesize p-doped samples, the investigation of the stability of the defects is necessary. The calculations done by Staab [14] and Kato et al. [15] tend to show that the most stable defects are multivacancies and interstitial magnesium atoms. Since the methodology of these works cannot be directly compared (they are different in many aspects), we found it necessary to do new calculations. In the case of $Mg_2Ge$, n doping was obtained for crystal growth under Mg rich conditions, while both p and n doping were obtained when starting from stoichiometric conditions [18-19]. Since defects play an important role in the doping mechanism of $Mg_2X$ compounds in view to improve their thermoelectric properties, it is important to have a correct description of the stability and of the formation energies of the different $Mg_2X$ based compounds and of their defects. The aim of this study is thus to use first principles calculations with different functionals in order to understand the relative stability of $Mg_2Si$ and $Mg_2Ge$ based compounds and of the hierarchy of some intrinsic defects like substitutionals, antisites and vacancies. Our results are compared with available experimental or numerical data in order to determine reliable quantities with error bars which can be useful in atomic interaction potentials developments or in thermodynamical calculations for phase diagrams investigations. The n and p type doping of these systems are then discussed using the shift of the Fermi level with respect to the top of the valence band.

Methods and calculation details are presented in section 2. The results and discussions concerning the stability of phases and the formation energies of defects are presented in section 3. In section 4 whe show the electronic properties of the materials with defects before drawing the major conclusions..

## 2. Computational details

First-principles calculations are performed using two different kinds of *ab initio* methods. In the first case, for relaxation calculations, we have used the scalar relativistic all-electron Blöchl's projector augmented-wave (PAW) method [20,21] within the local density approximation (LDA) or the generalized gradient approximation (GGA) as implemented in the highly-efficient Vienna Ab initio Simulation Package (VASP) [22,23]. For the GGA exchange-correlation we have used two functionals: the Perdew-Wang parameterization (PW91) [24] and the Perdew-Berke-Erzenhof parametrization (PBE) [25]. We have adopted the standard version of the PAW potentials for Mg, Si and Ge atoms. A plane-wave energy cutoff of 350 eV was held constant for all the calculations (we have controlled that a cut-off of 500 eV modifies the results by less than 1%). Brillouin zone integrations were performed using Monkhorst-Pack k-point meshes [26] and the Methfessel-Paxton technique [27] with a smearing parameter of 0.2 eV. For the pure compounds we have used a 21x21x21 mesh. All the calculations of point defects were performed using 2x2x2 cubic supercells in real space containing $N_{tot}$ = 96 particles. In that case we have used a 3x3x3 k-point mesh for the structural relaxation and the total energy was converged to less than $10^{-3}$ eV/atom.

The formation enthalpy of the $Mg_2X$ (X=Si,Ge) alloys in eV/atom can be calculated with the following equation [28]:

$$\Delta H(Mg_2X) = E(Mg_2X) - (N_{Mg}E(Mg)/N_{tot} + N_X\ E(X)/N_{tot}) \quad (1)$$

where $E(Mg_2X)$, $E$(Mg) and $E$(X) are the equilibrium first-principles calculated total energies (in eV/atom) of the corresponding $Mg_2X$ compound, of Mg with hcp structure and of X with diamond structure, respectively. $N_{Mg}$ is the number of magnesium atoms and $N_X$ the number of X atoms. We have calculated the formation energy both in stoichiometric conditions and in conditions rich and poor in magnesium. In the first case, the formation energy of a particular defect in $Mg_2X$ (X=Si,Ge) in eV/defect can be calculated from the following equation:

$$E_D = \frac{\Delta H_D(Mg_2X) - \Delta H_0(Mg_2X)}{x_D} \quad (2)$$

where $\Delta H_D(Mg_2X)$, $\Delta H_0(Mg_2X)$ and $x_D$ are respectively :

- the formation enthalpy calculated (in eV/atom) for the 2x2x2 supercell of the $Mg_2X$ compound containing the corresponding defect
- the formation enthalpy calculated (in eV/atom) for the 2x2x2 supercell of $Mg_2X$ compound without the defect
- the proportion of defects in the 2x2x2 supercell of $Mg_2X$.

The two first values are calculated using equation (1).

In the second case, the formation energy of a particular defect is calculated as follows [29]:

$$E_D = E(supercell\ with\ defect) + n\ \mu(Mg) + n\ \mu(X) + E(Mg_{64}X_{32}) \quad (3)$$

Where *E(supercell with defect)* is the energy of the supercell with a given defect, $E(Mg_{64}X_{32})$ is the energy the perfect supercell, $\mu(Mg)$ is the chemical potential of Mg and $\mu(X)$ is the chemical potential of X (Si or Ge). When the defect is a vacancy, n=-1. When the defect is an interstitial atom, n=1. In Mg rich conditions, $\mu(Mg) = E(Mg)$, where $E(Mg)$ is the energy of bulk stable Mg. In these conditions $\mu(Si) = \mu(Mg_2Si)-2\ E(Mg)$. In the same manner, for conditions poor in Mg, $\mu(Si) = E(Si)$, where $E(Si)$ is the energy of bulk stable Si. In this case $\mu(Si)$ = ½ ($\mu(Mg_2Si)$-$E(Si)$). When using these different expressions of the chemical potential in eq. (3), we can determine the energy of the different defects in conditions rich in Mg and poor in Mg. For the Brillouin zone integrations [30], we have used the Tetrahedron method for both the primitive cell and the supercell . In the first case, we have used a 21x21x21 k-mesh and in the second case an 11x11x11 k-mesh. To insure the quality of the calculations of the electronic structure and their robustness, especially in the case of the defects, we have also performed calculations using an all electron code. The computations were done using the framework of the full-potential augmented plane wave + local orbitals (FLAPW+lo) method [31, 32] within the Wien2k code [33]. The exchange and correlation contribution is calculated within the generalized gradient approximation (GGA) and the PBEsol parametrization [34]. The use of a different exchange-correlation functional permits us to show that the electronic properties associated with the defects are not dependent of some specific details such as the type of exchange-correlation functional used (see section 4). Two types of basis sets are introduced to solve the Kohn-Sham equations: atomic like orbitals inside muffin-tin spheres centered at the atomic sites and a plane wave expansion in the interstitial region. The convergence is controlled by the product $R_{MT}$ . $K_{max}$, where $R_{MT}$ is the smallest atomic muffin-tin sphere radius in the unit cell and $K_{max}$ is the largest reciprocal-lattice vector. In our calculations $R_{MT}$ is set equal to 2.0 a.u. for all atoms and all compounds or pure elements. A value of $R_{MT}$ . $K_{max}$ = 9 has been determined from specific test cases to ensure a convergence below the required accuracy and has been applied for all subsequent calculations. The k mesh is set to different values leading to equivalent densities of k points in the reciprocal space for the various unit cells considered. The core states are calculated within the spherical potential of the atomic spheres in each self-consistent-field (SCF) iteration by solving numerically the radial Dirac equation. For the valence states only scalar relativistic effects, the Darwin shift and the mass velocity term are considered and spin-orbit coupling is neglected. The total energy is converged to $10^{-4}$ Ryd.

## 3. Formation energy and stability

### 3.1 Phase stability of $Mg_2X$ (X=Si,Ge)

The structural properties and formation energy of both compounds have been computed using the two different exchange-correlation functionals as described in section 2 (see table 1). For the lattice parameters, the dispersion of the results obtained with GGA functionals is smaller than 0.5% and 1% for $Mg_2Si$ and $Mg_2Ge$ respectively. In order to compare the measured lattice parameters at T = 300 K with first-principle calculations at T = 0 K, the thermal expansion coefficient must be taken into account. From [35], the experimental linear thermal expansion coefficient is 1.5 $10^{-5}$ $K^{-1}$ for $Mg_2Ge$ and can be estimated to 1.35 $10^{-5}$ $K^{-1}$ for $Mg_2Si$. With this correction, the calculated lattice parameters transform to 6.314-6.364 Å for $Mg_2Si$ and 6.356-6.364 Å for $Mg_2Ge$ and the lattice constant is thus slightly overestimated by PBE and PW91 (+1%).

Concerning the formation energy, the range of the experimental results is quite large notably in the case of $Mg_2Si$ where they lie between -0.22 and -0.31 eV/at. in the literature [36,37]. The most recent experiments using accurate solution calorimetry and a mass spectrometry study of the thermal dissociation of $Mg_2Si$ indicate respectively a formation energy between -0.22 and -0.25 eV/at. with the most accurate experiments giving a formation energy of about -0.23 eV/at. [36]. If we compare our results with these last experimental values, we find that our theoretical values are between 2/3 and 3/4 of the experimental values. Such an underestimation from first principles calculations is frequently observed, notably for GGA calculations [38]. Much more sophisticated and time-consuming calculation techniques should be used to improve the agreement with the experiments [38], but this is beyond the scope of the present work

In the case of $Mg_2Ge$, the dispersion of the experimental results is about 13% with formation energies between -0.355 and -0.4 eV/at. [39,40], In the same manner, DFT calculations underestimate the formation enthalpy of $Mg_2Ge$ (-13% to -36%)..

If numerous calculations of the lattice parameters are available for the two compounds, only recently results concerning the formation energy became available [10-15], especially for $Mg_2Ge$ [10,12,13]. Concerning $Mg_2Si$, the formation energy was found to be about -0.173 to -0.189 eV/at. (if we except the Staab's results [14]), whereas it is about -0.236 to -0.254 eV/at. for $Mg_2Ge$. Therefore, this agrees with our results and confirms that the formation energy of $Mg_2Ge$ is about two third larger than the one of $Mg_2Si$.

Assessed values for the formation enthalpy of $Mg_2Si$ resulting from a fit to experimental data gives -0.218 eV/atom to -0.225 eV/atom [36,37,41,42] at 300 K. For $Mg_2Ge$, an assessed value of -0.362 eV/atom at 273 K was determined using the CALPHAD approach [39].

Using a double-zeta basis plus polarization (DZP) orbitals with the SIESTA code, Staab found a lattice parameter of 6.38 Å [14]. This value is relatively high with respect to calculations available in the literature or realized in the present work. The specified values for the formation enthalpies (calculated and measured) by Staab are surprising. Indeed, -0.0820 eV and -0.1178 eV per $Mg_2Si$ is found for the experimental and the calculated value respectively, which is definitively much smaller than experimental [36,37] and calculated values (present work and literature) using a plane wave basis [10-15]. Since, no reference about neither the exchange correlation functional nor the measurements are provided by the authors, these results must be taken with caution (see comments in Refs. [43] and [44]).

As a preliminary conclusion on the phase stability of $Mg_2X$ (X = Si, Ge), we can note that for $Mg_2Si$ and $Mg_2Ge$, our DFT calculations are in good agreement with the experiment: $Mg_2Ge$ has a slightly higher lattice parameter (+0.7%) and a higher formation enthalpy (about one half larger experimentally and about two-third larger in our calculations).

## 3.2 Formation energies of defects

### 3.2.1 Vacancy formation energy in pure materials

Prior to calculate the formation energies of defects in the $Mg_2X$ compounds, we calculate the formation energy of a vacancy in the pure elements: Mg, Si and Ge. The aim is to compare our results with those of other authors and to validate the methodology used in the rest of the paper. First, we discuss the case of a vacancy in semiconductors like silicon and germanium. As expected, we can see that for both compounds the lattice parameters are slightly larger than the experimental values (within 1 %) with the PBE and PW91 functionals. After full relaxation of the ionic positions and of the volume, we find respectively 3.61-3.65 and 2.14-2.2 eV/atom for the formation energy of a vacancy in the two materials. These values compare well with the values found in the literature for *ab initio* calculations. Indeed, in the case of GGA calculations, the values for Si lie between 3.1 and 3.81 eV /atom [45,46]; for Ge, the values lie between 2.56 and 2.6 eV/atom [46,47]. We notably find a very similar tendency than van Hellemont and coworkers who find respectively 3.81 and 2.56 eV/atom for Si and Ge [46] using also a PBE exchange-correlation functional. The comparison with experimental values is very favorable in the case of germanium where values close to 2.35 (±0.1) eV/atom have been estimated [47]. In the case of silicon, the comparison is more difficult due to the disagreement in the literature [45,48]. Indeed, based on EPR and DLTS experiments, Watkins proposes a value of 3.15 (-0.1-+0.3) eV/atom which is close to the GGA theoretical results [45], while in a discussion, Bracht proposes a value of about 4 eV/atom based on vacancy-mediated dopant diffusion experiments in silicon and hybrid functional calculations [48].

In the case of Mg, the lattice parameter is slightly smaller than the experimental value (within 1 %) in all cases and we find a much smaller value for the formation energy of a vacancy than in the elements above, namely 0.77-0.8 eV. This low value agrees well with the reported experiments ranging between 0.58 and 0.9 eV [49,50] and previous calculations (0.83+/-0.07 eV) [50] and is one of the reasons why the magnesium atoms diffuse very easily.

Despite the above discrepancies, which are related to the experimental difficulties to determine accurately the formation energy of vacancies as well as the approximations inherent to *ab initio* calculations, we still note that the comparison between experiments and theory is quite satisfactory. The *ab initio* PBE functional reproduces reasonably well the general experimental tendencies. We also notice that we find lower energy values than Staab: he found 0.97 eV for Mg and 4 eV for Si [14]. In the case of silicon, the situation is still unclear, in spite of a permanent refining of the experimental formation energy; in the case of Mg, the situation seems simpler and our results are very consistent with the experimental results whereas Staab's results [14] are larger than the experimental values [49,50].

### 3.2.2 Formation energies of defects in $Mg_2X$ (X = Si, Ge)

The formation energies of the different kinds of defects are reported in tables 3 and 4 for the $Mg_2Si$ and $Mg_2Ge$ compounds, respectively. We have assumed vacancies in the two different crystallographic sites, antisite defects and more complex defects such as an interstitial defect at the 4b Wyckoff position, bi- and tri-vacancies. In the case of $Mg_2Si$, only vacancies and antisite defects have been calculated by Imai et al [11]. Other kinds of defects such as interstitial defects at the 4b Wyckoff crystallographic site were studied by Kato et al [15] and multivacancies were studied by Staab [14]. Kato has calculated the defect in conditions rich and poor in magnesium. To our knowledge, no *ab initio* study has been dedicated to defects in $Mg_2Ge$.

In all cases, we can make the following observations:

- Vacancies on Si or Ge sites ($V^{Si}$ or $V^{Ge}$), antisite defects with Mg on Si or Ge sites ($Mg^{Si}$ or $Mg^{Ge}$) and interstitial Si or Ge atoms on 4b sites ($I^{Si}$ or $I^{Ge}$) are very unfavorable compared to other defects whatever the conditions are.

- Multivacancies ($V^{MgSi}$, $V^{Mg2Si}$), antisite defects with Si or Ge on Mg sites ($Si^{Mg}$ or $Ge^{Mg}$) and interstitial Mg atoms on 4b sites ($I^{Mg}$) are the most favorable defects compared to other defects whatever the conditions are.

In the case of $Mg_2Si$ in stoichiometric conditions, we find that the most favorable defects are interstitial Mg at the 4b site and multi-vacancies and the formation energy increases as:

$$V^{MgSi} < I^{Mg} < V^{Mg2Si} < V^{Mg} < Si^{Mg}$$

We find the same sequence in conditions rich in Mg, with the difference that $Si^{Mg}$ becomes very unstable and has even a smaller stability than $V^{Si}$.

In conditions poor in Mg, we find the following sequence:

$$V^{MgSi} < Si^{Mg} \sim< V^{Mg2Si} < I^{Mg} \sim< V^{Mg}$$

From these results, we can conclude that $V^{MgSi}$ is always the most stable defect, whatever the conditions are. In stoichiometric conditions and for conditions rich in Mg, we find that $I^{Mg}$ closely followed by the trivacancy $V^{Mg2Si}$ are the next most stable defects. On the contrary, in conditions poor in Mg, $I^{Mg}$ becomes a much less stable defect whereas the antisite defect $Si^{Mg}$ becomes as stable as the trivacancy.

Now, we discuss these results with respect to the literature.

In stoichiometric conditions, Imai et al. [11] found that the formation energy of the defects increases as:

$$V^{Mg} < V^{Si} \sim Si^{Mg} < Mg^{Si}$$

Note that he has not computed the case of interstial defects. Our results differ since we find that the $Si^{Mg}$ antisite defect has a similar formation energy than the $V^{Mg}$ vacancy while Imai et al. find a value two times larger for the first defect. This difference cannot be due to the exchange-correlation functional since we have used the same functional (PW91) than Imai et al. The discrepancies could arise from a different cut-off energy or a different pseudopotential. Indeed, we have tested that a cut-off energy of at least 300 eV is required to obtain well-converged results. In contrast, Imai et al. took a cut-off energy of 200 eV in their supercell calculations and used some corrections in order to decrease the errors induced by a too small cut-off [11] -this point remains questionable. Secondly, we have used PAW pseudopotentials while Imai et al. have used the less accurate norm-conservative pseudopotentials for the supercell calculations instead of the ultra-soft pseudopotentials [11] they have used for the primitive cell calculations. Using dual space Gaussian pseudopotentials and the LDA exchange-correlation functional, Kato et al [15] have calculated the formation energies of point defects in $Mg_2Si$ in conditions rich and poor in Mg. In the first case, these authors found that the Mg interstitial defect at the 4b site is always the most stable defect in $Mg_2Si$ considering both neutral and charged defects. In the case of conditions poor in Mg, they found that the most stable defect is $Si^{Mg}$, although we can see in their paper that, at least for the neutral case, the interstitial defect $I^{Mg}$ has formation energy that is comparable under these conditions. Comparing these results with ours is not straightforward since Kato et al considered charged defects with the formation energy depending on the Mg and Si chemical potentials. So we can compare our results with the ones of Kato et al only for the case of neutral defects. In that case, their results show the same tendency than the one observed in our calculations, except that we find a larger formation energy for the interstitial defect $I^{Mg}$ compared to $Si^{Mg}$ and $V^{Mg}$ in the case of conditions poor in Mg. However, Kato et al have not studied the case of multivacancies and our study indicates that it is necessary to do such a study in this class of materials because these defects have a low energy of formation and can be stabilized. Also, we note that their absolute values for the formation energies are smaller than ours. This could be explained by the use of the GGA functional instead of the LDA one, since the lattice parameters are systematically larger with the GGA description.

Although the calculations of Staab raise some questions (see section 3.1 and [43], [44]), we compare our results with his since it is the only other study of multivacancies in $Mg_2Si$. As discussed above, he has used a different type of local basis functions called DZP [14] provided in the SIESTA code and he has found larger formation energies for vacancies in Si and Mg than the ones obtained with a plane wave basis. We also notice that the estimates for the formation energies of vacancies are less consistent with the reported experimental values, most prominently in the case of Mg. Staab found that Si vacancies are more stable (E = 1.04 eV) than Mg vacancies (E = 1.74 eV) [14], a result which is in contradiction not only with our results and Imai et al's results [11] but also with Kato's work [15]. Indeed, the ratio between the energy of these two defects $V^{Si}/V^{Mg}$ is 0.6 [14], to compare with 1.44-1.47 in our case, 1.42 for Si-rich conditions and 0.95 for Mg-rich conditions in Kato's work [15] and 2.77 in Imai et al's work [11]. Even if the absolute values are not directly comparable, the results that the bi- and tri-vacancies have smaller formation energies than the monovacancies are confirmed in our work. This shows that multi-vacancies are among the most stable defects in $Mg_2Si$ and have to be considered as candidates for defects being present in this compound.

In summary, our calculations indicate that multivacancies are among the most stable defects in $Mg_2Si$ whatever the growth conditions are and more particularly the bivacancy is always the most stable defect. In conditions rich in Mg, the $I^{Mg}$ insterstitial defects are the most stable point defects and have stability comparable to multivacancies, in agreement with previous works. In conditions poor in Mg, the $Si^{Mg}$ antisite defect is the most stable point defect with a stability comparable to multivacancies. We find also that the $V^{Mg}$ and the $I^{Mg}$ defect (in that order) have slightly larger formation energies and should thus compete with these defects.

In the case of stoichiometric $Mg_2Ge$, the most favorable defects are Mg vacancies $V^{Mg}$, Ge on a Mg site and the trivacancy and the formation energy increases as

$V^{Mg} < V^{Mg2Ge} \sim Ge^{Mg} < V^{MgGe} < I^{Mg}$ in the case of the PBE functional

$V^{Mg2Ge} \sim V^{Mg} < V^{MgGe} < Ge^{Mg} < I^{Mg}$ in the case of the PW91 functional

In conditions rich in Mg, we obtain:

$I^{Mg} < V^{Mg2Ge} \sim V^{MgGe} < V^{Mg} < V^{Ge} < Ge^{Mg}$ in the case of the PBE functional

$V^{MgGe} < V^{Mg2Ge} < I^{Mg} < V^{Mg} < V^{Ge} < Ge^{Mg}$ in the case of the PW91 functional

In conditions poor in Mg, we obtain:

$Ge^{Mg} < V^{Mg} < V^{Mg2Ge} < V^{MgGe} < I^{Mg}$ in both cases

Because some defects have very close formation energies, it is difficult to say what the most stable defects are. However we can say that in stoichiometric conditions, the two most favorable defects are the $V^{Mg}$ vacancy and the $V^{Mg2Ge}$ tri-vacancy. The $V^{MgGe}$ bi-vacancy and the $Ge^{Mg}$ antisite defects are only slightly less stable and should enter in competition with the two other defects. In the case poor in Mg, the antisite defect becomes the most stable defect, whereas in growth conditions rich in Mg, the interstitial defect has a comparable stability with the multivacancies. From the above results for $Mg_2Si$ and $Mg_2Ge$, we can conclude that:

- In growth conditions rich in Mg: the multivacancies and the $I^{Mg}$ defects are the most stable defects.

- In stoichiometric conditions: the $I^{Mg}$ intertistial defect is clearly less favorable in $Mg_2Ge$ than in $Mg_2Si$ and, conversely, the $V^{Mg}$ vacancy is more stable in $Mg_2Ge$ than in $Mg_2Si$, whereas multivacancies are very favorable defects in both cases.

- In growth conditions poor in Mg: the antisite defect $Ge^{Mg}$ becomes the most stable defect just slightly below the multivacancies, whereas the antisite defect $Si^{Mg}$ has a stability comparable to the multivacancies in $Mg_2Si$. We note that $I^{Mg}$ becomes less stable in $Mg_2Ge$ than in $Mg_2Si$ compared to the other defects.

Finally, our main conclusion is that $I^{Mg}$ intertistial defect is clearly less favorable in $Mg_2Ge$ than in $Mg_2Si$ and that, conversely the $V^{Mg}$ vacancy and the $Ge^{Mg}$ becomes more stable in $Mg_2Ge$ than in $Mg_2Si$, whereas multivacancies are very favorable defects in both cases. Therefore, the defects related to lower Mg concentrations become more stable in $Mg_2Ge$ compared to the defects related to larger Mg concentration.

Now, we discuss the effect on the formation energies of the value of the bandgap that is underestimated in DFT calculations. We note that in the case of neutral vacancies in pure Si and Ge, this effect is small since the agreement of our results with the experiments is within 10 % (see section 3.2.1). Therefore, there is no reason that this effect is larger in the $Mg_2X$ compounds that have smaller bandgaps than pure Si and Ge. It is also worth noting that Kato et al. have also performed their calculations using DFT calculations (within the LDA) [15]. As can be seen from their paper, the underestimation of the bandgap has probably a larger effect in the case of charged defects than in the case of neutral defects.

Since these defects induce different kinds of doping, we will show in the next section how our results can explain the different types of electrical conductivities observed experimentally in $Mg_2Si$ and $Mg_2Ge$.

3.3 Electronic structure of pure and defects-containing $Mg_2X$ (X = Si, Ge)

Our calculations with VASP using the PBE parametrization for the pure compounds give an energy bandgap of 0.23 eV for $Mg_2Si$ and of 0.165 eV for $Mg_2Ge$. Using the all-electron Wien2k code and the PBEsol parametrization, we find $E_g$ = 0.1 eV in the case of $Mg_2Si$. This value is slightly underestimated by the broadening which is applied in order to integrate the DOS.

These results are in good agreement with other DFT based results in the literature: between 0.1-0.3 eV for $Mg_2Si$ [3,11,15,51,52] and between 0.1-0.16 eV for $Mg_2Ge$ [51,52]. The slightly larger energy band-gap found for the silicide agrees with the experimental tendency [1]. These values are lower than the experimental values (0.66-0.78 eV for $Mg_2Si$ and 0.57-0.74 eV for $Mg_2Ge$ [1,5]) as usually observed with standard DFT calculations, especially for narrow gap semiconductors, due to the fact that the excited states are not properly modeled within the standard DFT [5,53]. Methods based on the GW approximation are more adapted for accurate band structure investigations and confirm that the energy band gap is indirect and slightly larger in $Mg_2Si$ (0.65 eV) than in $Mg_2Ge$ (0.5 eV) [5]. However, calculations of the electronic DOS of a large supercell containing defects using this technique are not possible due to the large CPU time and memory requirements.

From Born effective charge calculations, Kato et al [15] have suggested that bondings in $Mg_2Si$ have a strong ionic character. In that case, one would expect for the most stable defects that :

- the interstitial magnesium $I^{Mg}$ and the divacancy $V^{MgSi}$ act as electron donors (n doping)
- the magnesium vacancy $V^{Mg}$ and the antisite $Si^{Mg}$ act as electron acceptors (p doping)
- the trivacancy $V^{Mg2Si}$ has a non doping effect.

For $Mg_2Si$, we have done DOS calculations for all these kinds of defects (not shown) by using the PBE exchange-correlation functional (VASP) and the PBEsol exchange-correlation functional with the all-electron Wien2k code. From these calculations, we determine the Fermi level shift induced by the defect and hence the doping. To do that, we define $\Delta E = E_F - E_V$, the difference between the Fermi level $E_F$ and the top of the valence band $E_V$. If $\Delta E < 0$, a p-doping is induced. If $\Delta E < 0$, an n-doping is induced. The results for the most stable defects are reported in table 5 and one can see that our calculations do not confirm the above naïve picture. The relatively good correspondence between both calculations [54] reinforces our confidence in such a conclusion and permits us to make the same conclusion for $Mg_2Ge$ for which only VASP calculations of the most stable defects have been done (see Table 6).

We note that for a long time, the nature of the bonding in $Mg_2X$ compounds is controversial in the literature. Indeed, the most recent calculations indicate that the bonds are partly covalent and partly ionic and that the charge on the Mg atoms is between +0.6 and +0.9 (depending on the type of calculation) instead of +2 as expected from the purely ionic picture [3]. However, analysis of the most recent XPS and X-ray fluorescence experiments indicate the almost covalent nature of the bonds in $Mg_2Si$ since its ionicity was estimated to be only of 8 %. The effective charge on Si was estimated to be -0.35 leading to a charge of 0.175 on Mg. [55,56]. We note also that recent NMR experiments agree much better with the covalent picture than with the ionic one [57]. Therefore, since the bonds have a significant covalent nature, it explains why the ionic description fails for the electronic properties of the intrinsic defects in both $Mg_2Si$ and $Mg_2Ge$. Although for the Mg vacancies and the Mg interstitial defects, we find the same behavior (see table 5) as the one predicted by the ionic picture, this is not the case for both the $X^{Mg}$ antisite defect and the multi-vacancies (see the different figures). Indeed, bivacancy defects have no doping effect, whereas the trivacancy defects act as electron acceptors, like the magnesium vacancy $V^{Mg}$ [58]. Also it appears that the vacancies have the surprising effect to increase the energy bandgap without doping, contrary to the naïve expectation of simple doping. This effect is stronger in the case of vacancies in the silicon sub-lattice and in the case of multivacancies. We note that in the case of the $V^{Mg}$, the Fermi level is shifted inside the valence band leading to two electrons less compared to the case of pure $Mg_2X$, whereas in the presence of a $V^{X}$, the Fermi level is shifted in the conduction band leading to two additional electrons compared to the case of pure $Mg_2X$. This observation permits to understand why there is no doping effect in the case of the bivacancy and why the trivacancy has the same doping effect than the $V^{Mg}$.

In the case of $Mg_2Ge$, the $Ge^{Mg}$ antisite defect has no doping effect and this is because the additional states introduced by the defect are inside the valence band and therefore the Fermi level stays inside the gap. In the case of $Mg_2Si$, the additional states introduced by the $Si^{Mg}$ antisite defect fill in the gap and the Fermi level moves at an energy slightly higher than the top of the valence band. However, since DFT calculations underestimate the value of the gap, we suggest that the $Si^{Mg}$ antisite has probably no doping effect. Note also that Kato et al have shown that the $Si^{Mg}$ antisite defect did not give any doping effect [15]. Since we find the same result for the $Ge^{Mg}$ antisite defect, this suggests that the $X^{Mg}$ antisite defects in both $Mg_2X$ compounds do not induce any doping. In all cases, we note that our results clearly contradict the naïve ionic picture.

In the case of interstitial $I^{Mg}$ defects, for both compounds, energy levels are introduced inside the energy gap in addition to induce n-doping since in that case the Fermi level shifts clearly inside the conduction band.

In our calculations the interstitial and antisite defects can fill up the energy gap, because of their large concentration, since we are using a 96 atoms supercell. However, in reality, the amount of these defects is small, and only few additional states are present in the gap. In that case, the main effect of the interstitial magnesium atoms is to shift the Fermi level in the conduction band whereas this is not the case for the $X^{Mg}$ antisite defect.

To conclude this part, only the interstitial Mg atoms can induce n-doping, whereas Mg vacancies and trivacancies give p-doping. The remaining other stable defects have no doping effect. In the next section, we discuss the consequences of these results for both compounds.

### 3.4 Discussion

As discussed in the introduction, whatever the growth conditions are, they systematically result in an n-type intrinsic doping for $Mg_2Si$ [16-18] contrary to the case of $Mg_2Ge$. Indeed, in this last case, it was found that both p and n doping were obtained when starting from stoichiometric conditions [18-19]. However, in Shank's study [19], the question of magnesium losses remained open when starting from stoichiometric conditions, but he showed in a convincing manner that intrinsic p doping is found when very pure starting elements are used (n doping occurs only when a significant Mg excess exists). These studies therefore suggest that defects giving p doping are much more stable in $Mg_2Ge$ than in $Mg_2Si$, whereas in both compounds, we can get n doping in conditions rich in magnesium. How can our calculations explain this behavior? We will discuss in detail each growth condition before drawing a more general conclusion.

In the case rich in Mg, we have seen that the three most stable defects are the bivacancies, the Mg insterstitial ($I^{Mg}$) and the trivacancies. Only the $I^{Mg}$ defect can give an n-doping whereas the trivacancies induce p-doping. We do not discuss the case of bivacancies since they have no doping effect. We note that in our calculations, the $I^{Mg}$ defect has a stability comparable to the trivacancies in $Mg_2Ge$ whereas in $Mg_2Si$ it is significantly more stable than the trivacancies.

This could explain why Shanks found that $Mg_2Ge$ becomes an n-type semiconductor only for a relatively large magnesium excess [19].

In the stoichiometric case, there are significant differences between the defect stability between $Mg_2Ge$ and $Mg_2Si$. If in the last case, the three most stable defects are still the bivacancies, the $I^{Mg}$ and the trivacancies, now the trivacancies have a similar stability than the $I^{Mg}$ and compete with it. However, comparison with experiments suggests that in reality the $I^{Mg}$ defects must have a larger stability compared to the trivacancies in stoichiometric conditions than what our calculations suggest. This is the only possibility to explain the experimental data. Concerning $Mg_2Ge$, we have found that the $V^{Mg}$ and $Ge^{Mg}$ defects become much more stable than in $Mg_2Si$ and even more stable than the $I^{Mg}$ defect with a stability comparable to the multivacancies. As $Ge^{Mg}$ has a neutral effect on doping, it is the $V^{Mg}$ and $V^{Mg2Ge}$ defects that give the p doping observed by Shanks in very pure samples grown under stoichiometric conditions [19].

In the conditions poor in Mg, there are also large differences in the defect stability between $Mg_2Ge$ and $Mg_2Si$. If in this last case, the antisite $Si^{Mg}$ defect has a comparable or better stability than the $I^{Mg}$ defect, it would have no impact on the doping as discussed in the previous section. Therefore, since the other stable defects inducing doping effects are the trivacancies and the $I^{Mg}$, the same remarks than in the stoichiometric situation still hold. In the case of $Mg_2Ge$, the antisite $Ge^{Mg}$ becomes a very stable defect but it has no impact on the doping. The $V^{Mg}$ becomes more stable than the trivacancies and must be at the origin of the p doping in that case. We can conclude that for both compounds, the case poor in Mg results in the same doping than in the stoichiometric case because the main change is the larger stability of the antisite $X^{Mg}$ defect which has no doping effect.

To summarize the results above, we have found that the interstitial $I^{Mg}$ defect is much less stable in $Mg_2Ge$ than in $Mg_2Si$ whereas the antisite $X^{Mg}$ and $V^{Mg}$ defects become more stable and this could explain why in $Mg_2Ge$ intrinsic p-doping is also possible in stoichiometric and poor Mg conditions. In this case, the p doping must be due to both the $V^{Mg}$ and the $V^{Mg2Ge}$ defects. As the interstitial defect $I^{Mg}$ is still very stable in conditions rich in Mg, this explains why $Mg_2Ge$ can be n-doped when the amount of magnesium excess is large enough. The high stability of this interstitial $I^{Mg}$ defect explains why $Mg_2Si$ is always of n type, although in stoichiometric and conditions poor in Mg, the $V^{Mg2Si}$ defects that can induce p doping could compete with it. However, as $Mg_2Si$ is always of n type this last defect must be less stable than the $I^{Mg}$.

Our calculations explain therefore quite naturally the experimental results on intrinsic defects in both $Mg_2Si$ and $Mg_2Ge$ and especially the results of Shanks concerning the last compound [19]. To conclude, it is because the defects inducing p-doping such as the $V^{Mg}$ defects and the trivacancies have a higher stability in $Mg_2Ge$ than in $Mg_2Si$ and because the defects inducing n-doping have a lower stability in $Mg_2Ge$ than in $Mg_2Si$ that p-doping becomes possible in $Mg_2Ge$ whereas it is not observed in $Mg_2Si$.

These results must be taken into account when these materials and their alloys are doped (and especially p-doped) in order to avoid undesirable compensation effects that could decrease the thermoelectric performance of these materials.

Finally we wish to note that during the review process Han and Shao published an article [59] on $Mg_2Si$ based tin alloys in which they showed also that the $I^{Mg}$ point defect is at the origin of the n-doping in pure $Mg_2Si$.

## 4. Conclusion

Using ab-initio calculations, we have studied the stability of point defects and multivacancies in the semiconducting $Mg_2X$ (with X = Si, Ge) compounds with the aim to understand the origin of their intrinsic doping.

Our study of the stability and electronic properties of point defects and multivacancies in $Mg_2X$ shows that the stability of the defects is strongly depending of the growth conditions as expected. In growth conditions rich in Mg, the multivacancies and the interstitial defects are the most stable defects. In stoichiometric conditions, when going from $Mg_2Si$ to $Mg_2Ge$, we find that the $I^{Mg}$ defect becomes less favorable, whereas the $V^{Mg}$ defect becomes more stable and the multivacancies are still very favorable. In growth conditions poor in Mg, the antisite $X^{Mg}$ defect becomes very stable and even the most stable defect in $Mg_2Ge$, with the multivacancies being also very favorable defects.

Since in both compounds the bivacancies and the antisite $X^{Mg}$ defects have no doping effect, our results imply that in poor Mg conditions, p doping is induced by $V^{Mg}$ and $V^{Mg2X}$ in both compounds. This should also be the case in stoichiometric conditions in the case of $Mg_2Ge$, whereas in conditions sufficiently rich in Mg, n doping is induced by the $I^{Mg}$ defect as in $Mg_2Si$. For this last compound, this defect can induce n-type doping also in other growth conditions, although the trivacancies that induce p-type doping, have a comparable stability in conditions rich in Si. Since in experiments $Mg_2Si$ is always of n type, the most stable defect has to be the $I^{Mg}$ defect. All these results agree qualitatively with experimental results in which a systematic n-type doping is found in $Mg_2Si$, whereas in $Mg_2Ge$ the doping can be of p or n type depending on the growth conditions.

Finally, our main conclusion is that the $I^{Mg}$ intertistial defect is clearly less favorable in $Mg_2Ge$ than in $Mg_2Si$ and that, conversely the $V^{Mg}$ vacancy and the $Ge^{Mg}$ become more stable in $Mg_2Ge$ than in $Mg_2Si$, whereas multivacancies are very favorable defects in both cases. Therefore, the defects related to lower Mg concentrations become more stable in $Mg_2Ge$ compared to the defects related to larger Mg concentration and this explains naturally the difference in intrinsic doping between the two compounds.

These results have consequences that have to be taken into account when these compounds are intentionally doped. Indeed since different defects are present in $Mg_2Si$ and $Mg_2Ge$, they must have some impact not only on the induced doping (i. e. some compensating effects can appear) but also on the solubility limits of the doping impurity. This can have some significant impact on the optimization of the thermoelectric properties of these materials which are systematically doped for such applications.

**Acknowledgements**

The authors thank the “Programme Interdisciplinaire Energie” (PIE) of the CNRS for financial support. Most of the calculations have been performed on the CINES and the HPC@LR computer centers.

**Table captions :**

**Table 1:** The calculated lattice parameters and formation enthalpies for $Mg_2Si$ and $Mg_2Ge$. a : ref. 1 ; b : refs. 36, 37 ; c : refs. 39, 40

**Table 2:** The calculated lattice constants and vacancy formation energies for silicon, germanium and magnesium. a: refs.45, 46, 48; b : refs.46, 47; c : refs. 49, 50.

**Table 3:** The calculated defect formation energies (in eV/atom) for the different types of defects in $Mg_2Si$.

**Table 4:** The calculated defect formation energies (in eV/atom) for the different types of defects in $Mg_2Ge$.

**Table 5:** The difference between the Fermi level and the top of the valence band: $\Delta E = E_F - E_V$ and the corresponding doping for the most stable defects ($E_{form} < 2$ eV) in $Mg_2Si$ (with decreasing stability in stoichiometric conditions). The case of the $V^{Si}$ vacancy is also given since it becomes more stable in conditions poor in Mg. In parenthesis the results from the All-electrons calculations (AE) are also given.

**Table 6:** The difference between the Fermi level and the top of the valence band : $\Delta E = E_F - E_V$ and the corresponding doping for the most stable defects ($E_{form} < 2$ eV) in $Mg_2Ge$ (with decreasing stability in stoichiometric conditions). The case of the $V^{Ge}$ vacancy is also given since it becomes more stable in conditions poor in Mg.

**Table 1**

| Phase | Calculation type or experiment | Structure type | Lattice parameters a(Å) | Formation enthalpy (eV/atom) |
|---|---|---|---|---|
| $Mg_2Si$ | GGA-PBE (PAW) | $CaF_2$ | 6.356 | -0.162 |
| | GGA-PW91 (PAW) | | 6.356 | -0.186 |
| | Experiment | | 6.34-6.39[a] | -(0.22-0.31)[b] |
| $Mg_2Ge$ | GGA-PBE (PAW) | | 6.4228 | -0.277 |
| | GGA-PW91 (PAW) | | 6.419 | -0.306 |
| | Experiment | $CaF_2$ | 6.385-6.393[a] | -(0.355-0.4)[c] |

**Table 2**

| Phase | Calculation type or experiment | Lattice parameters of stoichiometric compound a(Å) | c(Å) | Vacancy formation enthalpy (eV/vacancy) |
|---|---|---|---|---|
| Si | GGA-PBE (PAW) | 5.4684 | 5.4684 | 3.65 |
| | GGA-PW91 (PAW) | 5.4676 | 5.4676 | 3.61 |
| | Experiment | 5.43 | 5.43 | 3.2-4[a] |
| Ge | GGA-PBE (PAW) | 5.781 | 5.781 | 2.1 |
| | GGA-PW91 (PAW) | 5.7792 | 5.7792 | 2.2 |
| | Experiment | 5.6575 | 5.6575 | 2.35±0.1[b] |
| Mg | GGA-PBE (PAW) | 3.1927 | 5.173 | 0.77 |
| | GGA-PW91 (PAW) | 3.1994 | 5.16 | 0.8 |
| | Experiment | 3.21 | 5.211 | 0.58-0.9[c] |

**Table 3**

| Defects in $Mg_2Si$ | PW91 Mg-rich side | PW91 stoichio | PW91 Si-rich side | PBE Mg-rich side | PBE stoichio | PBE Si-rich side |
|---|---|---|---|---|---|---|
| $V^{Mg}$ | 1.72 | 1.54 | 1.44 | 1.68 | 1.53 | 1.44 |
| $V^{Si}$ | 1.89 | 2.26 | 2.44 | 1.89 | 2.21 | 2.37 |
| $Si^{Mg}$ | 2.22 | 1.66 | 1.39 | 2.04 | 1.55 | 1.32 |
| $Mg^{Si}$ | 2.27 | 2.83 | 3.10 | 2.24 | 2.83 | 2.98 |
| $I^{Si}$ at 4b site | 2.94 | 2.57 | 2.38 | 2.72 | 2.39 | 2.24 |
| $I^{Mg}$ at 4b site | 1.21 | 1.40 | 1.49 | 1.11 | 1.26 | 1.35 |
| $V^{MgSi}$ | 1.07 | 1.16 | 1.21 | 1.07 | 1.15 | 1.19 |
| $V^{Mg2Si}$ | 1.42 | 1.42 | 1.42 | 1.33 | 1.33 | 1.33 |

**Table 4**

| Defects in $Mg_2Ge$ | PW91 Mg-rich side | PW91 stoichio | PW91 Ge-rich side | PBE Mg-rich side | PBE stoichio | PBE Ge-rich Side |
|---|---|---|---|---|---|---|
| $V^{Mg}$ | 1.46 | 1.16 | 1.00 | 1.44 | 1.17 | 1.03 |
| $V^{Ge}$ | 1.96 | 2.52 | 2.82 | 1.93 | 2.49 | 2.76 |
| $Ge^{Mg}$ | 2.16 | 1.3 | 0.84 | 2.07 | 1.25 | 0.83 |
| $Mg^{Ge}$ | 2.17 | 3.04 | 3.49 | 2.13 | 2.96 | 3.37 |
| $I^{Ge}$ at 4b site | 2.79 | 2.23 | 1.93 | 2.65 | 2.1 | 1.82 |
| $I^{Mg}$ at 4b site | 1.32 | 1.62 | 1.77 | 1.22 | 1.49 | 1.63 |
| $V^{MgGe}$ | 1.07 | 1.20 | 1.28 | 1.25 | 1.39 | 1.46 |
| $V^{Mg2Ge}$ | 1.14 | 1.14 | 1.14 | 1.24 | 1.24 | 1.24 |

**Table 5**

| Compound | Defect (decreasing stability) | Fermi level shift [eV] PAW-PBE (AE) | Doping type |
|---|---|---|---|
| $Mg_2Si$ | $V^{MgSi}$ | 0 (0) | No |
| | $I^{Mg}$ | 0.45 (0.25) | n |
| | $V^{Mg2Si}$ | -0.15 (-0.20) | p |
| | $V^{Mg}$ | -0.20 (-0.25) | p |
| | $Si^{Mg}$ | 0.25 (0.20) | No (see text) |
| | $V^{Si}$ | 0.60 (0.50) | n |

**Table 6**

| Compound | Defect (decreasing stability) | Fermi level shift [eV] PAW-PBE | Doping type |
|---|---|---|---|
| $Mg_2Ge$ | $V^{Mg}$ | -0.20 | p |
| | $V^{Mg2Ge}$ | -0.15 | p |
| | $Ge^{Mg}$ | 0 | No |
| | $V^{MgGe}$ | 0 | No |
| | $I^{Mg}$ | 0.30 | n |
| | $V^{Ge}$ | 0.55 | n |